\documentclass{jaa}
\usepackage{natbib}
\usepackage[colorlinks=true,citecolor=blue]{hyperref}
\usepackage{comment}

\usepackage{graphicx}
\usepackage{amsmath}
\usepackage{gensymb}

\begin{document}\sloppy

\title{Investigating star formation activity in the Sh 2-61 H\,{\sc ii} region}


\author{Rakesh Pandey\textsuperscript{1,2}, Saurabh Sharma\textsuperscript{1}, Lokesh Dewangan\textsuperscript{2}, Aayushi Verma\textsuperscript{1}, Tapas Baug\textsuperscript{3}, Harmeen Kaur\textsuperscript{4} and Arpan Ghosh\textsuperscript{1} }
\affilOne{\textsuperscript{1}Aryabhatta Research Institute of Observational Sciences (ARIES), Manora Peak, Nainital 263 002, India.\\}
\affilTwo{\textsuperscript{2}Physical Research Laboratory (PRL), Navrangpura, Ahmadabad - 380 009, India.\\}
\affilThree{\textsuperscript{3}Satyendra Nath Bose National Centre for Basic Sciences (SNBNCBS), Block-JD, Sector-III, Salt Lake, Kolkata-700 106, India\\}
\affilFour{\textsuperscript{4}Center of Advanced Study, Department of Physics DSB Campus, Kumaun University Nainital, 263002, India\\}


\twocolumn[{
\maketitle

\corres{pandey.rakesh405@gmail.com}

\msinfo{15 November 2022}{15 November 2022}

\begin{abstract}

Using the multiwavelength data sets, we studied the star formation activity in H\,{\sc ii} region Sh 2-61 (hereafter S61).
We identified a clustering in the region and estimated the membership using the Gaia proper motion data. 
The physical environment of S61 is inspected using infrared to radio wavelength images. We also determined the Lyman continuum flux associated with the H\,{\sc ii} region and found that the  H\,{\sc ii} region is formed by at least two massive stars (S1 and S2). We also analyzed the $^{12}$CO(J =3$-$2) JCMT data of S61, and a shell structure accompanying three molecular clumps are observed towards S61. We found that the ionized gas in S61 is surrounded by dust and a molecular shell. Many young stellar objects and three molecular clumps are observed at the interface of the ionized gas and the surrounding gas. The pressure at the interface is higher than in a typical cool molecular cloud.
\end{abstract}
 

}]


\doinum{12.3456/s78910-011-012-3}
\artcitid{\#\#\#\#}
\volnum{000}
\year{0000}
\pgrange{1--}
\setcounter{page}{1}
\lp{1}

\section{\label{In} Introduction} 

Massive stars play a crucial role in the evolution of their host galaxies through their stellar winds, strong ultra-violet radiation, and supernova explosions. 
The strong feedback from the massive stars deeply affects the natal molecular cloud by changing the physical condition of the cloud. 
The massive stars, which are often seen associated with the young star cluster,  H\,{\sc ii} regions in a typical star-forming region, 
can increase or decrease the star formation rate in their surroundings. 
The increase and decrease in the star formation rate are termed as the positive and negative feedback of the massive star, respectively. 
The relative impact of the positive and negative feedback process depends upon the immediate environment around the
 massive star itself \citep{2017MNRAS.467..512S}.
However, the formation of the second generation of stars due to the positive feedback of massive stars 
is reported in the literature \citep{2005A&A...433..565D,2022ApJ...926...25P}, but still lack statistically significant sample to conclude on the theoretical models responsible for it.

\begin{figure*} 
\centering
\includegraphics[width=0.9\textwidth]{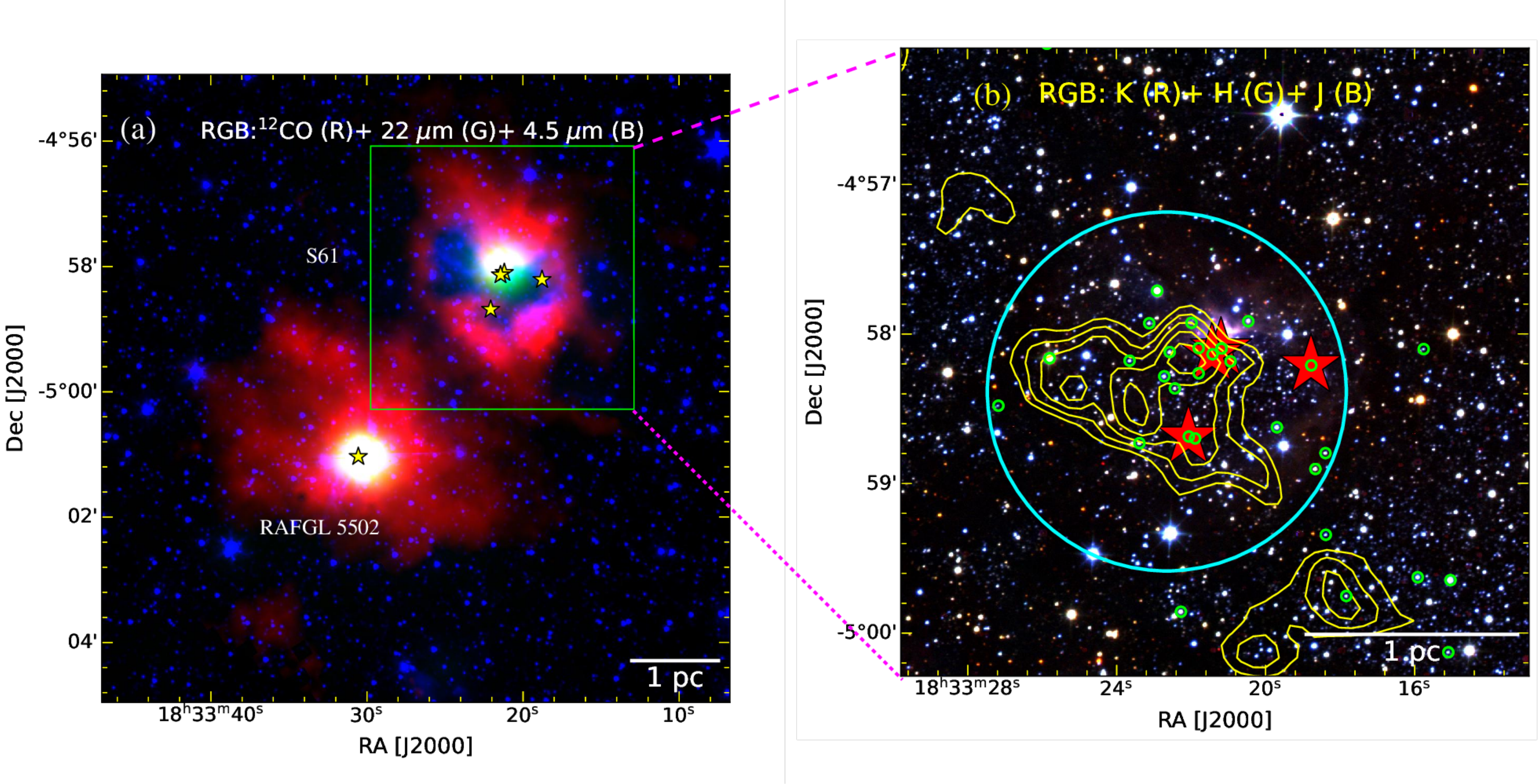}
\caption{\label{fi1} (a) Color-composite image covering a large FOV ($\sim$10$^{\prime}$$\times$10$^{\prime}$) around our studied target. The image is made using $^{12}$CO moment-0 map (red), WISE 22 $\mu$m (green), 4.5 $\mu$m (blue) band images. The massive stars inside the FOV are marked with star symbols, and the target region studied in the present work (S61) is marked with a green box. (b) Color-composite image of S61 (FOV $\sim$ 4.2$^{\prime}$$\times$4.2$^{\prime}$) made using UKIDSS K (red), H (green), J (blue) band images. The massive stars in S61 are marked with star symbols, while small green circles denote the member stars.
A large circle of radius 1$^{\prime}$.2 enclosing the identified clustering, massive stars, and most of the member stars are shown with cyan color.
}
\end{figure*}

The target for the present study is the galactic  H\,{\sc ii} region Sh 2-61 (hereafter S61; $\alpha$$_{2000}$ =18$^{h}$33$^{m}$20$^{s}$, $\delta$$_{2000}$ = -4$\degree$58$^{\prime}$0$^{\prime \prime}$,
FOV $\sim$ 4.2$^{\prime}$$\times$4.2$^{\prime}$). The star-forming complex Avedisova 437 \citep{2002ARep...46..193A}, hosting the target region, lies above the galactic plane (b=1.68$\degree$) in the constellation of Aquila. The cloud hosting the target site is shown in Figure \ref{fi1}(a), which extends from north-west to south-east direction; it also hosts a young massive star-forming region rafgl 5502 \citep{2002ARep...46..193A}. The clump accommodating rafgl 5502 hosts a 6.7 GHz methanol maser \citep{2019ApJS..241...18Y}, a massive protostar \citep{2013MNRAS.430.1125C}, while S61 seems comparatively older, hosting evolved massive stars \citep{1990AJ.....99..846H,2004AJ....127.1682H}. The target region S61 hosts a Herbig Ae/Be star AS310 (hereafter S1; $\alpha$$_{2000}$ =18$^{h}$33$^{m}$21.9$^{s}$, $\delta$$_{2000}$ = -4$\degree$58$^{\prime}$05.9$^{\prime \prime}$) \citep{2004AJ....127.1682H} along with B type massive stars S61-2 (hereafter S2, $\alpha$$_{2000}$ =18$^{h}$33$^{m}$21.44$^{s}$, $\delta$$_{2000}$=-4$\degree$58$^{\prime}$08.2$^{\prime \prime}$), S61-3 (hereafter S3; $\alpha$$_{2000}$ =18$^{h}$33$^{m}$22.07$^{s}$, $\delta$$_{2000}$=-4$\degree$58$^{\prime}$41.3$^{\prime \prime}$), and S61-4 (hereafter S4; $\alpha$$_{2000}$ =18$^{h}$33$^{m}$18.78$^{s}$, $\delta$$_{2000}$=-4$\degree$58$^{\prime}$12.7$^{\prime \prime}$) \citep{1990AJ.....99..846H}. S61 was initially misclassified as a planetary nebula now known to be an H\,{\sc ii} region \citep{2013MNRAS.431....2F, 2014ApJS..212....1A}, ionized by at least three B type stars \citep{1990AJ.....99..846H}, the region is also reported in the  H\,{\sc ii} region catalog of \citet{2014ApJS..212....1A}. Most of the previous studies on the target region \citep[e.g.,][]{2006ApJ...653..657M, 2018A&A...620A.128V} are mainly focused on the Herbig star S1 such as determining mass, age, luminosity, and studying emission line activity of the star.
In Figure \ref{fi1}(a), we have shown a color composite image of a large region (FOV $\sim$10$^{\prime}$$\times$10$^{\prime}$), showing the morphology of the cloud hosting the target site S61. The
image is made using the  $^{12}$CO moment-0 map (red), WISE 22 $\mu$m (green), and Spitzer 4.5 $\mu$m (blue) image. In the image, our target site is enclosed within a box showing morphology such as an embedded cluster, massive
stars, and signature of a feedback-driven structure. Despite this interesting morphology, no previous effort has been made in determining the cluster parameters, studying dust and gas distribution, cloud morphology, census of
young stellar objets (YSOs), the role of stellar feedback, and star formation activity in the region. Therefore, we have studied this region using recently available high quality multiwavelength archival data sets (cf. Table \ref{archilog}).

The structure of this paper is as followings. The morphological parameters are estimated in Section \ref{cl} and Section \ref{mem}. The physical environment of the  H\,{\sc ii} region is examined using the multiwavelength data sets in Section \ref{ph}. In Section \ref{sf}, we discussed the star formation processes in S61 and concluded in Section \ref{con}.   

\section{Data sets} \label{sec2}
The archival data sets used in the present study are tabulated in Table \ref{archilog}. We made our near-infrared (NIR) catalog by obtaining the JHK magnitudes of all the sources inside the S61 region from the UKIDSS Galactic Plane Survey (GPS) \footnote{http://wsa.roe.ac.uk:8080/wsa/}. We applied the selection criteria of considering only those sources which have H and K band uncertainties less than 0.1 mag.

\section{Stellar clustering in the S61 \label{cl}} 

We have performed the stellar surface density analysis in this region to identify the grouping/clustering of stars. 
The surface density distribution of the stars is estimated using the nearest neighbor (NN) method, which has been described in our previous publications \citep{2020ApJ...891...81P,2022ApJ...926...25P}. In this method, we determine the local density of stars in each position of a uniform grid by measuring the projected radial distance to the Nth nearest neighboring star. The projected radial distance can be used to determine the local density value at each grid point \citep [cf.][]{2005ApJ...632..397G}. In the present work, we used a grid size of 10$^{\prime\prime}$ and varied radial distance to accommodate  20$^{th}$ NN in the UKIDSS (cf. Table \ref{archilog}) NIR photometric catalog. The resultant density distribution of the stars is shown as yellow contours in Figure \ref{fi1}(b). The lowest density contour is 1$\sigma$ (6 star/arcmin$^{2}$)
above the mean stellar density (28 star/arcmin$^{2}$) and the step size is 2 star/arcmin$^{2}$. 
A star cluster (hereafter `61cluster') of elongated morphology is clearly identified in this region. 
We have identified the active region (undergoing star formation) of S61 as a circle centered at $\alpha_{2000}$: 18$^{h}$33$^{m}$22$^{s}$.721, $\delta_{J2000}$: -4$\degree$58$^{\prime}$20.36$^{\prime \prime}$ with a radius of
\textbf{1$^{\prime}$.2} (shown with a cyan circle in Figure \ref{fi1}(b)) which encloses the identified cluster `61cluster', massive stars (see Section \ref{In}), and 
most of the member stars of S61 (see Section \ref{mem}). The member stars (green circles) and the active region (cyan circle) are shown in Figure \ref{fi1}(b).

\begin{figure*} 
\includegraphics[width=1\columnwidth]{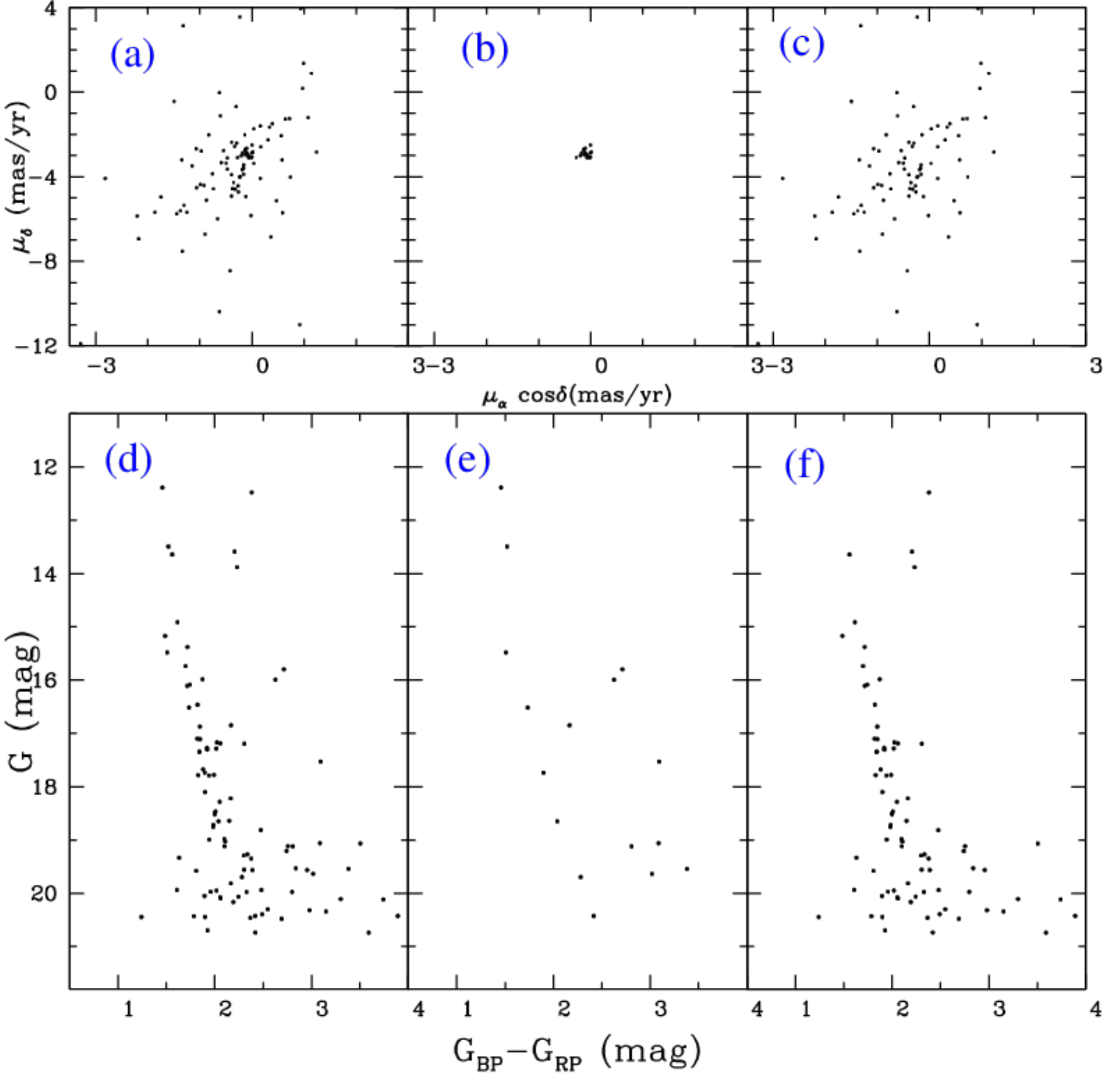}
\includegraphics[width=1\columnwidth]{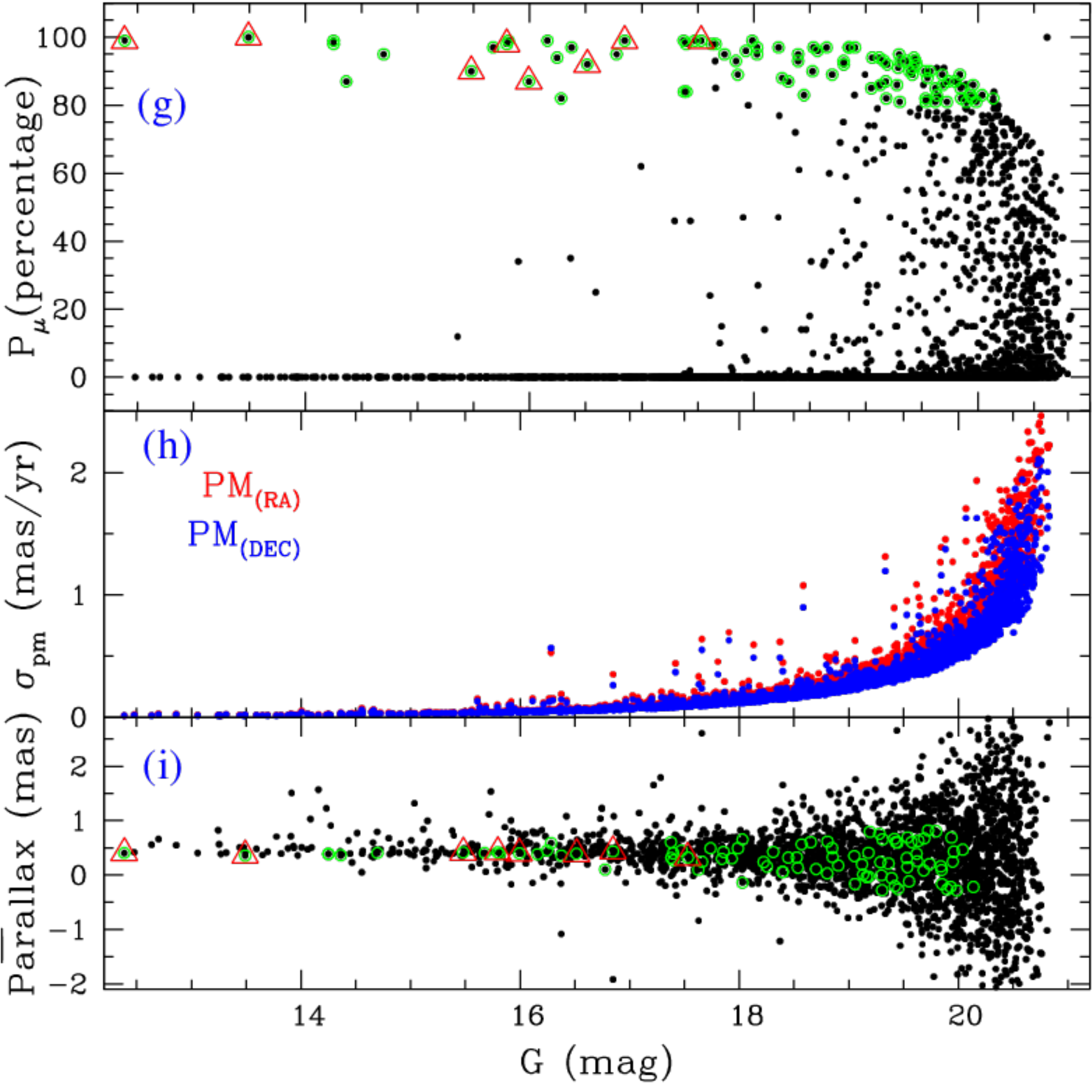}
\caption{\label{g3} Left panel: Vector point diagrams (a,b,c) and CMD (d,e,f) for all the stars (a,d), 
probable member stars (b,e), and field stars  (c,f).
Right panel: The membership probability (g), PM errors (h), and parallaxes (i) 
are plotted as a function of the $G$ magnitude. The green circles denote the member stars identified in
the present analysis (see Section \ref{mem}), and red triangles are the eight stars with very accurate parallax values (see Section \ref{mem}). 
 } 
\end{figure*}

\section{ \label{mem} Membership and Distance estimation } 

We have used the proper motion (PM) data of the recently released Gaia DR3 \citep{2021A&A...649A...1G} to identify stars that 
belongs to S61. In panel (a, b, c) of Figure \ref{g3}, we have plotted PM components $\mu_\delta$, and $\mu_\alpha$cos($\delta$) in the form of a vector point diagram (VPD). We considerd all the stars within the active region (within the cyan circle drawn in Figure \ref{fi1}(b)) which have have PM error less than 3 for plotting VPD diagram. Panel (a) shows the VPD for all the stars, while panel (b) and panel (c) show VPDs for the most probable stars and field stars, respectively. The most probable member stars are chosen visually by selecting stars inside a 0.6 mas yr$^{-1}$ radius around the
centroid of the cluster in PM space ($\mu_\alpha$cos($\delta$) = -0.22 mas yr$^{-1}$ and $\mu_\delta$ = -2.91 mas yr$^{-1}$) (panel (b)). Panel (d, e, f) of the Figure \ref{g3} shows CMDs for all the stars (d), for the probable member stars (e), and the field stars (f). 
The selection of the most probable member star is always a comprise between selecting the stars sharing the mean PMs but are field stars and neglecting the stars whose PMs are poorly defined but are member stars. We obtained the mean value of PMs and the corresponding dispersion for the member star as ($\mu_\alpha$cos($\delta$) = -0.22 mas yr$^{-1}$, $\mu_\delta$ = -2.91 mas yr$^{-1}$, $\sigma_c$, $\sim$0.5 mas yr$^{-1}$) and for the field star as (($\mu_{xf}$ = -0.67 mas yr$^{-1}$, $\mu_{yf}$ = -3.54 mas yr$^{-1}$, $\sigma_{xf}$ = 2.94 mas yr$^{-1}$ and $\sigma_{yf}$ = 3.60 mas yr$^{-1}$). These values are further used to generate the frequency distribution function and estimate the membership probability. The method is used and discussed in \citet[]{2020MNRAS.498.2309S,2020ApJ...891...81P}.

The membership probability is the ratio of the distribution of the cluster star to all the stars, given by the below-mentioned equation. 

\begin{equation}
P_\mu(i) = {{n_c\times\phi^\nu_c(i)}\over{n_c\times\phi^\nu_c(i)+n_f\times\phi^\nu_f(i)}}
\end{equation}

In the above equation, ($\phi^\nu_c$) and ($\phi^\nu_f$) represent the frequency distribution function for the cluster and field stars, respectively. $n_c$ (=0.17) and $n_f$ (=0.83) denote the normalized number of stars in cluster and field region, respectively. Based on above numbers we have calculated the membership probability for all the 
stars in our selected region of S61, i.e., $\sim$ 4.2$^{\prime}$$\times$4.2$^{\prime}$ (cf. Figure \ref{fi1} (b)).
The right-panel of Figure \ref{g3} shows the distribution of membership probability (g), error in PMs (h), and parallax (i) against the $G$ mag. We adopted the membership criteria of considering only those stars as members with membership probability 
higher than 80 \% (shown with green circles in panel (g) and panel (i)), which gives us 18 member stars in S61. This method effectively separates the member stars in the brighter magnitude ends but is not that effective at the fainter end (see panel (g, h, i) of Figure \ref{g3}). The reason behind this discrepancy is the higher uncertainties in PMs as we go towards the fainter magnitude ends.

Few distance estimates of this region are available in the literature, \citet{1990AJ.....99..846H} and
\citet{1990AJ....100.1915H} reported the distance as 2.3 kpc, which is determined by taking mean of the kinematic and photometric distance of the early type stars. The other distance estimate is 3.3 kpc by \citet{1984ApJ...279..125F}, which is calculated using CO velocity of the associated molecular cloud and CO rotation curve of the galaxy.
We have also calculated the distance of `61cluster' using the parallax of the member stars. We found that the out of the 18 member stars of S61, 
eight stars have very accurate parallax values (i.e., error$<$ 0.1 mas, red triangles in panel (i) of Figure \ref{g3}) 
and have distance estimated by \citet{2021AJ....161..147B}. 
We took the mean distance value of these member stars and constrained the cluster's distance as 2.4$\pm$0.2 kpc.

\begin{figure*}
\centering\includegraphics[height=18cm, width=18cm]{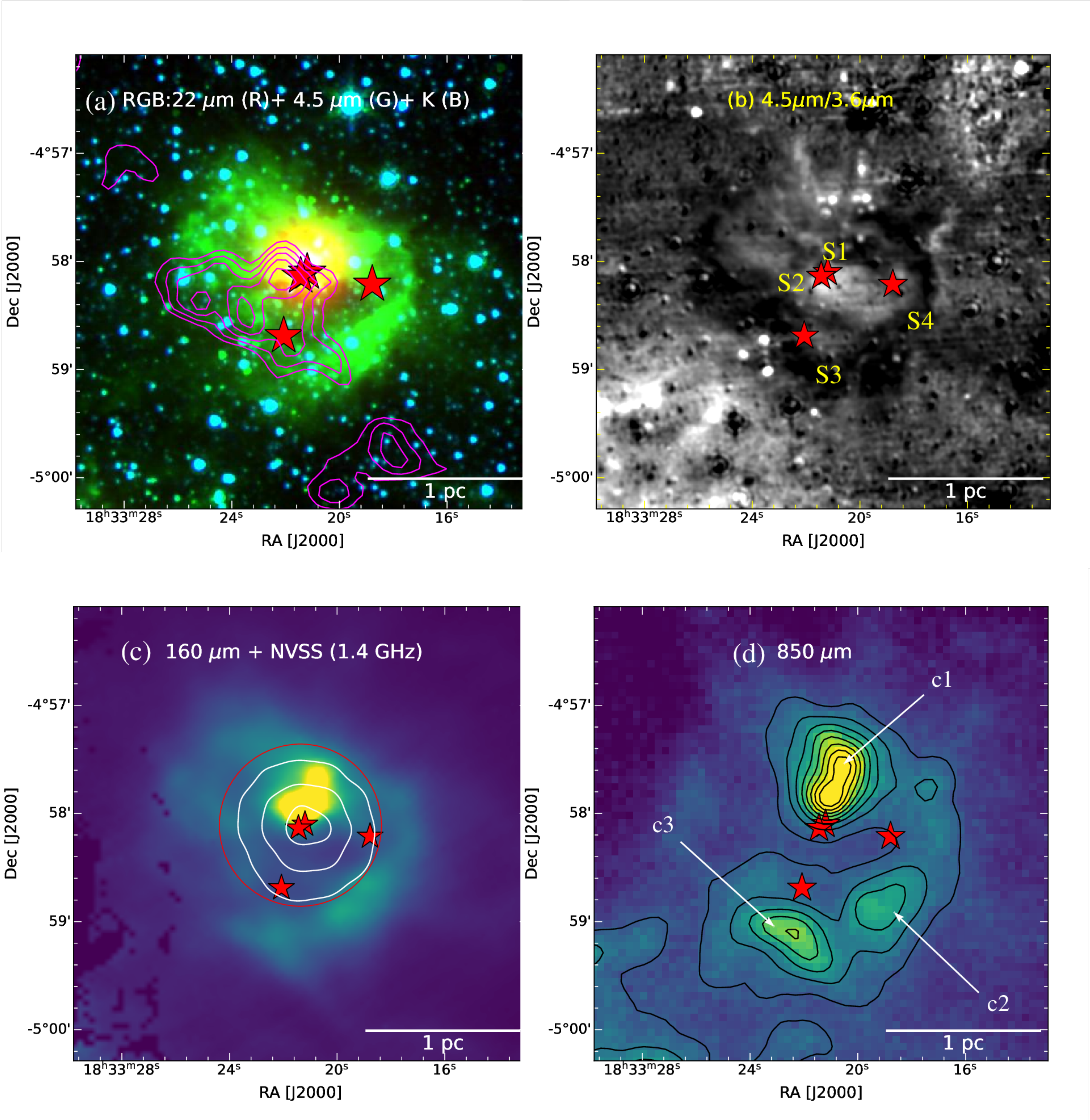}
\caption{\label{fig1}(a): Color composite image of S61 made using WISE 22 $\mu$m (red), Spitzer 4.5 $\mu$m (green), and NIR K band (blue images), the surface density contours (see Section \ref{cl}) are also overplotted on this image with magenta color. (b) Spitzer ratio map (4.5 $\mu$m/3.6 $\mu$m) of the region. (c) Herschel 160 $\mu$m overplotted with the NVSS radio contours (white color), a circle of radius 45$^{\prime \prime}$ centered at massive star S1, and S2 (cf. Section \ref{feedb}), is also shown with red color. The lowest NVSS contour is at 0.03 Jy/beam and the step size is 0.03 Jy/beam. (d). JCMT SCUBA-2 (850 $\mu$m) image of S61 is also overplotted with its own contours, the lowest contour is at 0.7 mJy/(arcsec)$^{2}$ and the step size is 0.3 mJy/(arcsec)$^{2}$. Three prominent clumps (c1, c2, and c3) are also marked in the image. The star symbols in all the images show the massive stars in S61. All the images cover the same FOV as in Figure \ref{fi1} (a).}
\end{figure*}

\section{Physical environment of the S61 \label{ph}} 

\subsection{Morphology of the S61 \label{du}}

The physical environment of S61 is traced by investigating the multiwavelength images spanning from NIR to
sub-mm wavelengths. In Figure \ref{fig1}(a), we show a color-composite image of the region in a FOV of 4.2$ \times$ 4.2 arcmin$^{2}$.
The image is made by using the $WISE$ 22 $\mu$m (red), $Spitzer$ 4.5 $\mu$m (green), and K band (blue) images and is also overplotted with the surface density contours (see Section \ref{cl}). This image shows a dust lane (traced using 4.5 $\mu$m emission) extended from the northeast to the southwest direction.
The clustering identified in the present work extends similarly and is bounded by this dust lane from the west direction.
A dust clump harboring the massive stars (S1 and S2) lies at the center of the extended dust lane where the dust envelope in the WISE 22 $\mu$m emission can be seen.

We have also produced the Spitzer ratio map (4.5 $\mu$m/3.6 $\mu$m) of this region, which is shown in Figure \ref{fig1}(b). The technique used for making the Spitzer ratio map and its effectiveness is discussed in \citet{2017ApJ...834...22D}. The Spitzer 3.6 $\mu$m band accommodates the prominent poly-cyclic aromatic hydrocarbon (PAH) band at 3.3 $\mu$m. In comparison, 4.5 $\mu$m band is a useful tracer of the shocked hydrogen gas, hosting several $H_2$ line
transitions \citep{2009ApJ...695L.120Y}, and also the Br-$\alpha$ emission (at 4.05 $\mu$m). The dark region in the ratio map tracing the PDR (photo-dissociation regions) region \citep[see for more details;][]{2017ApJ...834...22D} is mainly distributed in a ring-like structure around the massive stars having bright patches near them.
The PDR structure shows that the gas/dust surrounding the massive stars is getting significantly heated by them. The bright diffused region at the location of the massive stars, also accompanied by the radio continuum emission (see Figure \ref{fig1}(c)), indicates the Br-$\alpha$ emission due to the photoionization.
Note that a bright patch in the bottom left corner of the image does not belong to S61.
The few other bright spots not accompanied by radio continuum emission favor the ($H_2$) line emission and could be due to the outflow activity of young stars.

Figure \ref{fig1}(c) shows the color image of the region made using the Herschel mid-infrared (MIR) 160 $\mu$m image.
The Herschel image shows a cold dust clump near the position of the massive stars S1 and S2, and the NVSS radio continuum contours are centered on them.

We also show the JCMT SCUBA-2 850 $\mu$m image (see Table \ref{archilog}) of the region in Figure \ref{fig1}(d).
The image is also overplotted with its brightness contours. The 850 $\mu$m emission shows three prominent clumps with otherwise diffuse emission in the region.
The most prominent clump (c1) accommodates the massive stars S1 and S2, while the other two clumps (c2 and c3) are seen in the south direction of the massive stars S4 and S3, respectively.

\begin{figure}[h]
\includegraphics[width=1\columnwidth]{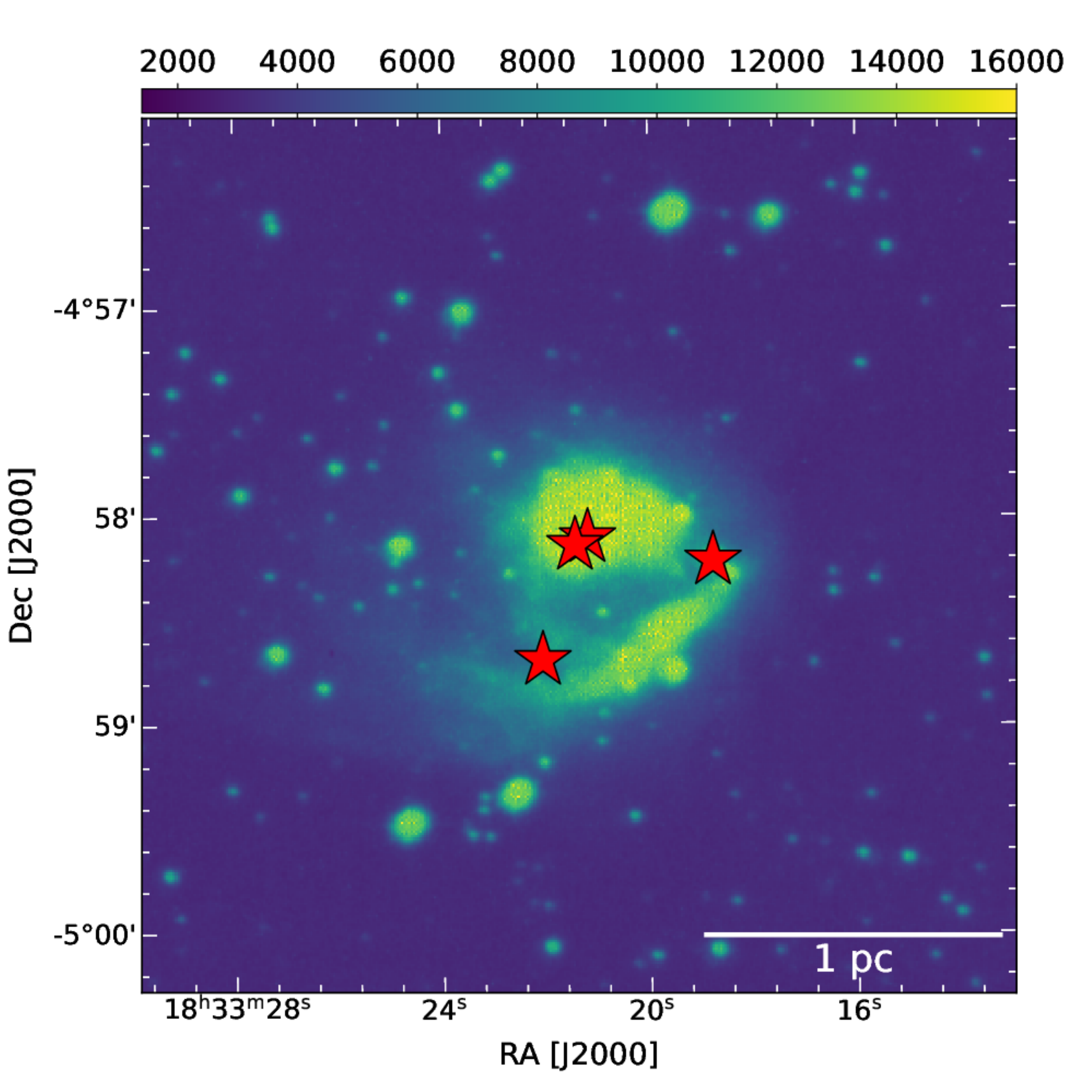}
\caption{\label{hal}The H$\alpha$ image of S61, the symbols have the same meaning as in Figure \ref{fig1}.}
\end{figure}

\subsection{Ionized gas in the S61  \label{io}}

The morphology of the ionized gas in S61 is traced using the NVSS radio continuum image (1420 MHz) along with the H$\alpha$ image. 
If we look at the NVSS radio continuum contours overplotted over the Herschel 160 $\mu$m image (see Figure \ref{fig1} (c)), 
the radio emission looks concentrated towards the central clump (c1) encircling the massive stars S1 and S2. The massive stars S1 or S2 or both
seem to be the ionizing source/sources responsible for creating this  H\,{\sc ii} region. The other two massive sources (S3 and S4) are located at the boundary of this  H\,{\sc ii} region.  
The H$\alpha$ emission, as seen in Figure \ref{hal}, is predominantly distributed towards the clump c1. 
An arc structure, probably separated from the main clump c1 by a dust lane, is also seen towards the south of massive stars S4 and S3. 

We calculated the Lyman continuum flux associated with the  H\,{\sc ii} region, which can be further used to constrain the spectral of the ionizing source/sources. The equation used for calculation is expressed as follows \citep{2016A&A...588A.143S}:

\begin{equation}
\begin{split}
N_\mathrm{UV} (s^{-1})& = 7.5\, \times\, 10^{46}\, \left(\frac{S_\mathrm(\nu)}{\mathrm{Jy}}\right)\left(\frac{D}{\mathrm{kpc}}\right)^{2} \\
&\left(\frac{T_{e}}{10^{4}\mathrm{K}}\right)^{-0.45} \times\,\left(\frac{\nu}{\mathrm{GHz}}\right)^{0.1}
\end{split}
\end{equation} 

In the above equation, N$_{UV}$ denotes the Lyman continuum photons per second, T$_e$ denotes the electron temperature (taken as 10,000 K), $\nu$ is 
the frequency, S$_\nu$ is the integrated flux, D is the distance of the  H\,{\sc ii} region (taken as 2.4 kpc, see Section \ref{mem}). This calculation assumes that a single massive source generates the ionization flux.
We applied the task `jmfit' of AIPS on the NVSS (1.4 GHz) map to estimate the integrated flux. The values of the integrated flux for S61 is found to be 0.17$\pm$.002 Jy ($\sigma$=0.0018 Jy/beam), 
and the diameter of the  H\,{\sc ii} region is found to be $\sim$ 79.5$^{\prime\prime}$ (0.9 pc). The obtained value of N$_{UV}$ is found to be 46.88 \citep[between B0.5V-B0V spectral type;][]{1973AJ.....78..929P}, which is greater than the combined Nuv of S1 and S2 \citep[two B1 type of stars, combined N$_{UV}$=45.59;][]{1973AJ.....78..929P}, and the radio continuum emission is peaking at their positions. The H\,{\sc ii} region seems to be formed by at least these two massive stars (S1 and S2).

The dynamical age of S61 is obtained by using the equation given by \citet{1980Natur.287..373D}, which is expressed as follows:

\begin{equation}
t_{dyn} = \Big(\frac{4R_s}{7c_s}\Big) [\Big(\frac{R_{H II}}{R_s}\Big)^{7/4}-1]
\end{equation}

where, c$_s$ is the isothermal sound velocity in the ionized gas (c$_s$ = 11 km s$^{-1}$) \citep{2005fost.book.....S}, R$_{H II}$ is the radius of the H\,{\sc ii} region, and R$_s$ is the  Str\"{o}mgren radius of the H\,{\sc ii} region, which is given by:

\begin{equation}
R_s = \Big(\frac{3S_{\nu}}{4\pi{{n_0}^2}{\beta_2}}\Big)^{1/3}
\end{equation}

where, n$_0$ is the initial ambient density (in cm$^{-3}$) and $\beta$$_2$ 
is the total recombination coefficient to the first excited state of hydrogen $\beta$$_2$ = $2.6\times10^{-13}$ \citep{2005fost.book.....S}.
We have estimated the the dynamical age of S61 as 0.2(0.6) Myr for n$_0$=10$^3$(10$^4$) cm$^{-3}$.

\subsection{Molecular gas kinematics in S61}

\begin{figure}
\includegraphics[width=1\columnwidth]{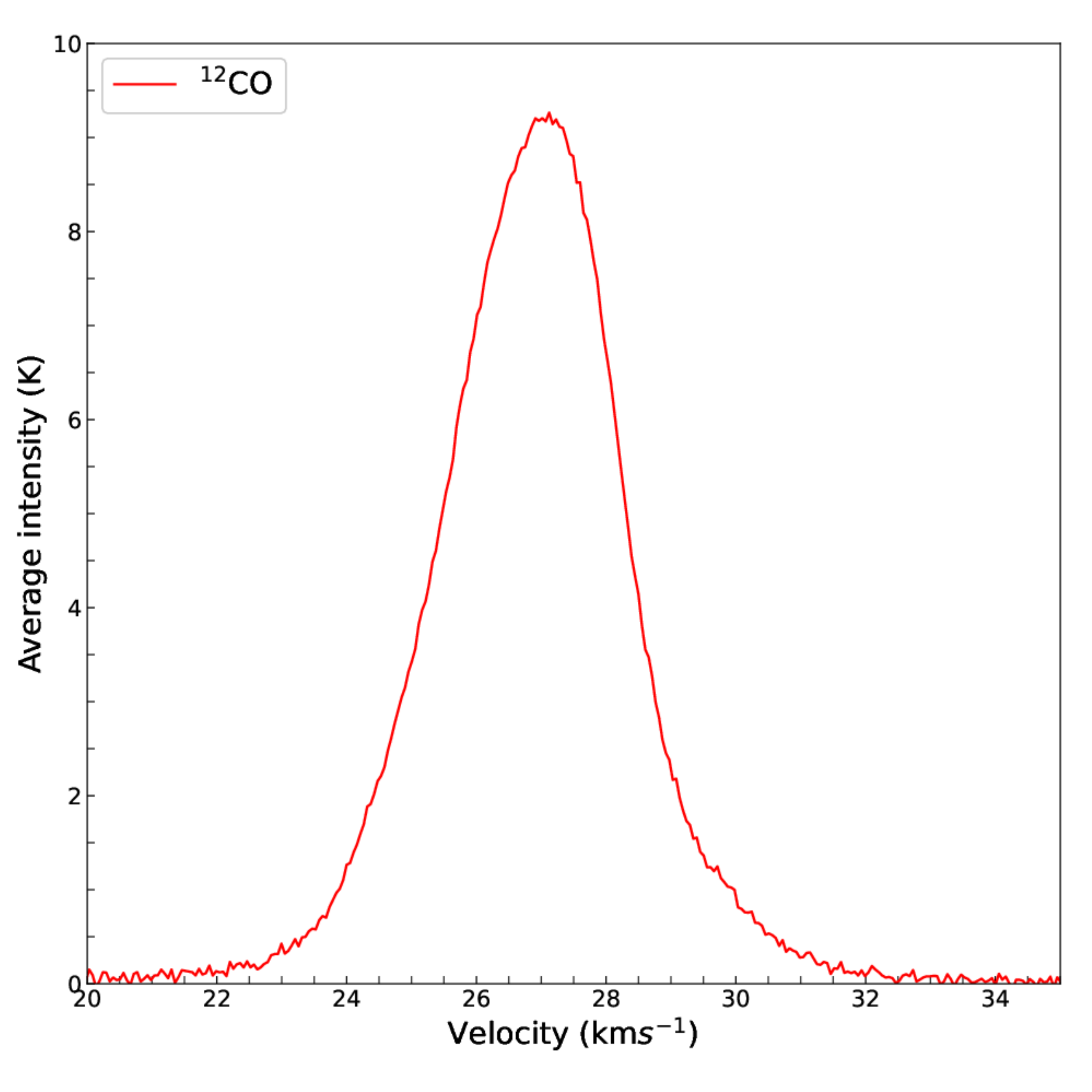}
\caption{\label{spec}The velocity profile of the $^{12}$CO(J =3$-$2) towards S61.}
\end{figure}

We traced the distribution of the molecular gas and investigated the gas kinematics in S61 using the $^{12}$CO(J =3$-$2) $JCMT$ data. 
The extracted velocity profile of the $^{12}$CO from S61 is shown in Figure \ref{spec}. The molecular gas in S61 is traced in a velocity range of 20-34 kms$^{-1}$.

\begin{figure*} 
\includegraphics[height=9cm,width=9cm]{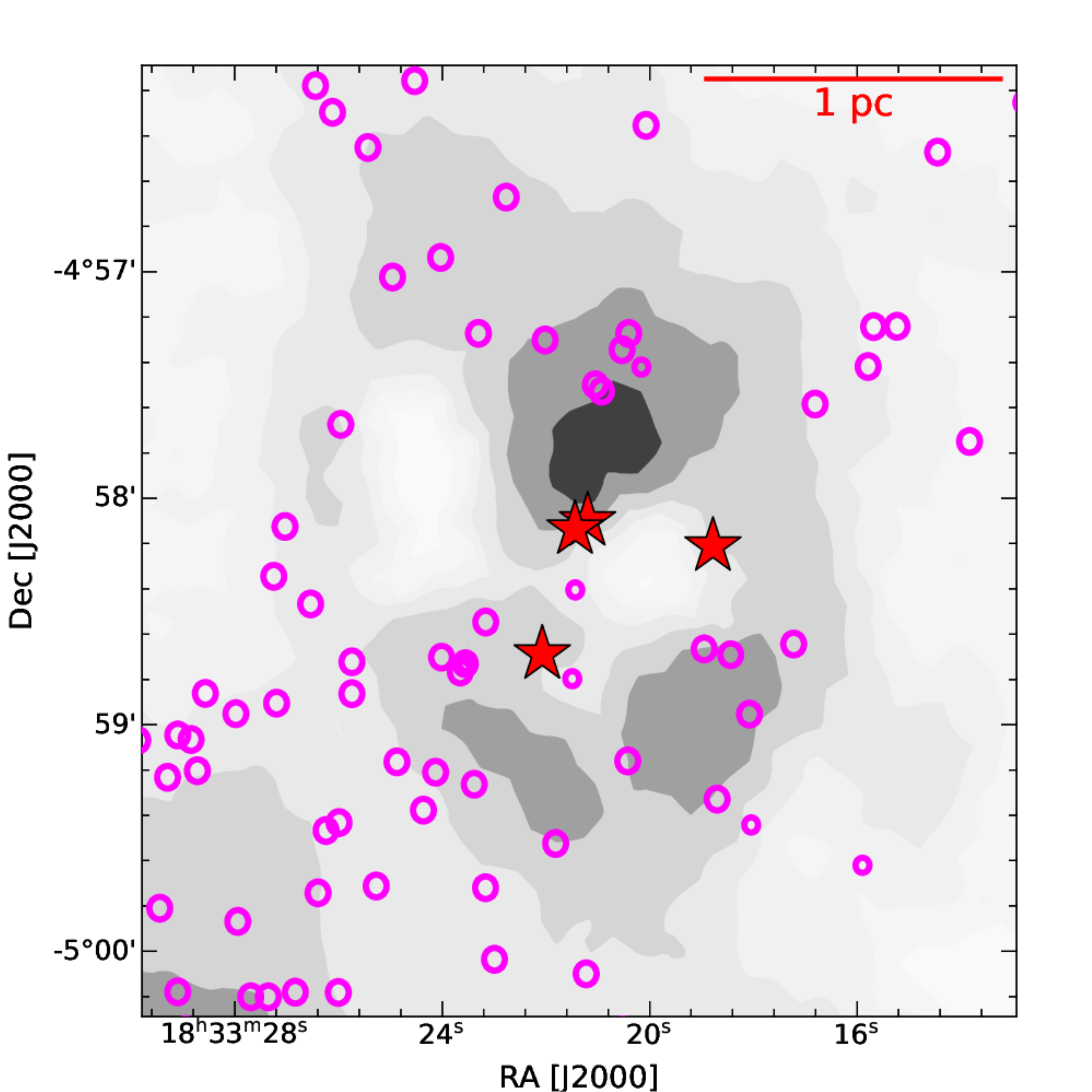}
\includegraphics[width=9cm,height=9cm]{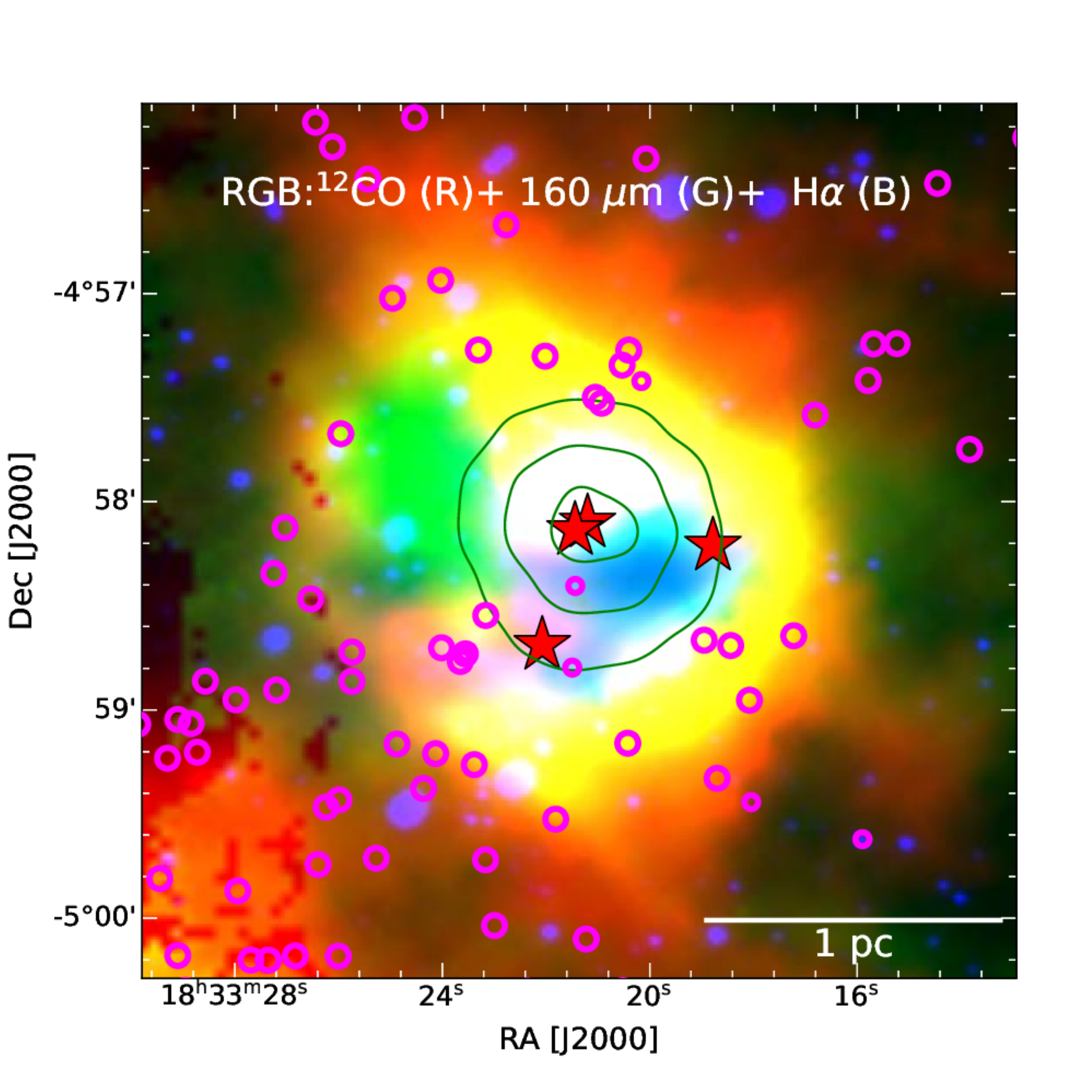}
\caption{\label{lst} (a) Velocity integrated $^{12}$CO map (20-34 kms$^{-1}$) of S61. (b) We show a color-composite image of S61 made using $^{12}$CO moment-0 map (red), Herschel 160 $\mu$m image, and H$\alpha$ (blue) images. The NVSS radio continuum contours are shown with the green color. The lowest NVSS contour is at 0.03 Jy/beam and the step size is 0.03 Jy/beam. In both figures, YSOs identified in the present study are marked with the magenta circles.}
\end{figure*}

In Figure \ref{lst}(a), we show $^{12}$CO intensity map of S61. A shell structure is visible on the map, accommodating the three prominent molecular clumps. 
We have also previously seen the dust clumps at these positions (see Section \ref{du}). 
The massive stars S1 and S2 lies at the inner edge of the clump c1, while massive stars S4 and S3 are situated nearby the clump c2 and c3, respectively.

\subsection{Pressure calculation at the boundary of S61} \label{feedb}

We also examined the effect of feedback from the massive stars in S61 by calculating the total feedback pressure. The total feedback pressure 
consists of the stellar wind pressure (P$_{wind}$) \citep{2012ApJ...758L..28B,2017ApJ...851..140D}, radiation pressure (P$_{rad}$), and the H\,{\sc ii} region pressure (P$_{ H\,{\sc ii}}$), which are given by the following equations \citep[see for details,][]{2012ApJ...758L..28B}.


\begin{equation}
P_{rad} = L_{bol}/ 4\pi c D_{s}^2; 
\end{equation}

\begin{equation}
P_{ H\,{\sc ii}} = \mu m_{H} c_{s}^2\, \left(\sqrt{3N_{UV}\over 4\pi\,\alpha_{B}\, D_{s}^3}\right);   
\end{equation}

\begin{equation}
P_{wind} = \dot{M}_{w} V_{w} / 4 \pi D_{s}^2; 
\end{equation}

In the above-mentioned equations, L$_{bol}$ denotes the bolometric luminosity of the ionizing source, $D_{s}$ is the distance at which pressure has to be calculated,  m$_{H}$ is the hydrogen atom mass,
c$_{s}$ is the sound speed in the photoionized region \citep[=11 km s$^{-1}$;][]{2009A&A...497..649B}. N$_{UV}$  in the above equation denotes the Lyman continuum flux, `$\alpha_{B}$' is the radiative recombination coefficient, and $\dot{M}_{w}$,  V$_{w}$ is the mass-loss rate, and the wind velocity of the ionizing source, respectively. 

In Section \ref{io}, we found that the spectral type of the ionizing source of this  H\,{\sc ii} region has a spectral type between B0.5V-B0V,
and the massive stars S1 and S2 are responsible for creating this  H\,{\sc ii} region. 
We calculated the total pressure at the boundary of S61 where molecular clumps and YSOs (cf. Appendix A) are observed. 
We considered B0.5V as the spectral type of the ionizing source and average distance of the molecular clumps 
from the massive central stars as 45$^{\prime \prime}$ (1.1 pc at a distance of 2.4 kpc). A circle of radius 45$^{\prime \prime}$ centered at massive central stars is shown in Figure \ref{fig1}(c). For a B0.5V star, we adopted the values of $L_{bol}$= 19952 L$_{\odot}$, $\dot{M}_{w}$ $\approx$ 2.5$\times$ 10$^{-9}$ M$_{\odot}$ yr$^{-1}$,  V$_{w}$ $\approx$ 1000 km s$^{-1}$ , and N$_{UV}$ = 2.9 $\times$ 10$^{47}$ \citep[]{2017ApJ...834...22D}. The value of the  $P_{ H\,{\sc ii}}$, $P_{rad}$, $P_{wind}$ is estimated as 1.19 $\times$ 10$^{-10}$ dynes\, cm$^{-2}$, 1.89 $\times$ 10$^{-11}$ dynes\, cm$^{-2}$, and 1.16 $\times$ 10$^{-13}$ dynes\, cm$^{-2}$, respectively. The total pressure ($P$= $P_{ H\,{\sc ii}}$+$P_{rad}$+ $P_{wind}$) value is estimated as 1.38$\times$ 10$^{-10}$ dynes\, cm$^{-2}$.

\section{Star formation activity in the S61\label{sf} }

The recent star formation activity in S61 are traced by massive stars, cluster of stars, YSOs, and the molecular clumps. In this section, we have discussed the implications of the observed morphology and investigated the processes of star formation in S61.

In Figure \ref{lst} (b), we show the color-composite image of S61 made by using the $^{12}$CO intensity map (red), Herschel 160 $\mu$m (green), and  H$\alpha$ (green) images. The figure shows a cometary morphology where the H$\alpha$ emission is seen to be surrounded by the 160 $\mu$m and molecular emission.
This morphology suggests that the interaction of the ionized gas (traced by the H$\alpha$ emission and radio continuum contours) with the surrounding dust and gas has possibly formed this structure. This kind of 
morphology, where the dust/gas emission encloses the ionized gas emission, is also quite common in the galactic bubbles \citep{2006ApJ...649..759C}. We have also seen three prominent molecular clumps at the interface of the ionized gas and the surrounding gas, suggesting that the gas's compression from the ionization front could have formed these clumps. The ratio map (shown in Figure \ref{fig1}) also shows the PAH emissions at the interface of the ionized emission and the surrounding gas, strongly suggesting the impact of the  H\,{\sc ii} region. We also see few YSOs distributed towards the molecular clumps where the ionized gas and the surrounding cloud interact (cf.  Figure \ref{lst} (a)). Most of these sources are Class II with only 2 Class I sources, which could have formed due to the ionizing feedback from the massive stars. This kind of morphology, where the feedback from the massive star triggers the star formation in the surrounding molecular cloud, is observed in various  H\,{\sc ii} regions such as Sh 2-301 \citep{2022ApJ...926...25P}, Sh 2-217 \citep{2011A&A...527A..62B}, Sh 2-212 \citep{2008A&A...482..585D}, and RCW 120 \citep{2010A&A...518L..81Z}. We also compared the average age of the YSOs with the dynamical age of S61. The estimated dynamical age (0.2(0.6) Myr) of the  H\,{\sc ii} region (considering ambient density of n$_0$=10$^3$(10$^4$) cm$^{-3}$) does not seems to be high enough to
trigger the formation of the YSOs which are considered to have an average age of 0.46 and 1-3 Myr \citep{2009ApJS..181..321E}. The total pressure at the boundary of the  H\,{\sc ii} region where the molecular clumps and YSOs are distributed is found to be 1.38$\times$ 10$^{-10}$ dynes\, cm$^{-2}$ (see Section \ref{feedb}). The obtained pressure value is one magnitude greater than the pressure of a cool molecular cloud (they considered to have the pressure of $P_{MC}$$\sim$10$^{-11}$--10$^{-12}$ dyne cm$^{-2}$ at a temperature of $\sim$20 K 
and particle density $\sim$10$^{3}$--10$^{4}$ cm$^{-3}$, Table 7.3 of \citet[][]{1980pim..book.....D}). 
The higher pressure value could be due to the
feedback from the massive stars that could have compressed the surrounding molecular cloud.
The overpressure at the boundary of the H\,{\sc ii} region and morphological suggestions are two observational signposts of feedback triggered star formation. The other signposts, such as the difference between the age of young stars and the dynamical age of the feedback agent, age gradient among the age of young stars, and elongation of young star cluster towards the feedback agent, are common in the feedback triggered star formation sites \citep{2015MNRAS.450.1199D}. A simultaneous occurrence of many signposts mentioned above can only be put as suggestive evidence for the feedback triggered star formation. Since there are processes like cloud-cloud collision \citep{2021PASJ...73S.405F}, and accretion through filaments \citep{2018ApJ...852..119B} that act at a larger scale, and can trigger star formation. We also need to inspect the larger area around the target region to explore the possibility of other processes acting at a larger scale before making any conclusion regarding feedback triggered star formation, which is going to be a prospect for future studies in the region.

\section{Conclusion\label{con}}

In this paper, we studied S61 with the multiwavelength data sets. The conclusions drawn from the study are as follows.

\begin{itemize}

\item
We identified a stellar clustering in S61 using the UKIDSS NIR data. From the population of member stars, the  distance to this region is estimated as 2.4$\pm$0.2 kpc.

\item
The region's dust distribution is examined using the Spitzer and Herschel archive images. The dust is found to be distributed in a shell kind of structure 
enclosing the H$\alpha$ emission and ionized gas. We also found PDR structure distributed in a ring-like structure around the massive stars (S1 and S2), showing heating of the dust.

\item
We traced the morphology of the ionized gas in S61 using the NVSS radio continuum data and H$\alpha$ image. The ionized gas is found to be distributed in a circularly symmetric way, peaking at the position of the massive stars S1 and S2. We also calculated the Lyman continuum flux and dynamical age of the   H\,{\sc ii} region, which are found to be 46.88 and (0.2-0.6) Myr, respectively.
We found that the spectral type of the ionizing source is between B0.5V-B0V, and the massive sources S1 and S2 might be collectively responsible for creating the  H\,{\sc ii} region.

\item
We also examined the kinematics of the molecular gas in S61 using $^{12}$CO(J =3$-$2) $JCMT$ data. A molecular shell accompanying the three molecular clumps is observed towards S61.

\item 
The dust and molecular gas in S61 are found to be surrounding the ionized gas and the massive stars. We also have significant PAH emissions at the interface of the ionized gas and the dust/molecular gas. The YSOs and molecular clumps are found to be distributed at the interface.

\end{itemize}

\appendix

\begin{table*} 
\centering
\caption{\label{archilog} Details of the archival data set used in the present study}
\footnotesize
\begin{tabular}{@{}lccccc@{}}
\hline
\hline
Survey & Details \\
      &       &  \\
\hline
\textcolor{red}{\bf Optical} \\
The AAO/UKST SuperCOSMOS H$\alpha$ survey  & \citep[SHS; resolution $\sim$1-2$''$;][]{2005MNRAS.362..689P}  \\
\textcolor{red} {\bf NIR} \\
United Kingdom Infrared Telescope Infrared Deep Sky Survey & \citep[UKIDSS; resolution $\sim$0.8$''$;][]{2008MNRAS.391..136L} \\
\textcolor{red} {\bf MIR} \\
Warm-{\it Spitzer} GLIMPSE360 Survey    & \citep[$\lambda$ =3.6 and 4.5 $\mu$m; resolution $\sim$2$''$;][]{2005ApJ...630L.149B} \\
Wide-field Infrared Survey Explorer & \citep[WISE, $\lambda$ =22 $\mu$m; resolution $\sim$12$''$;][] {2010AJ....140.1868W}  \\
\textcolor{red} {\bf FIR} \\
Herschel infrared Galactic Plane Survey (Hi-Gal) & \citep[$\lambda$ = 160 $\mu$m; resolution$\sim$12$''$;][]{2010PASP..122..314M}\\ 
\textcolor{red} {\bf Sub-millimeter} \\
JCMT SCUBA-2 Guaranteed Time projects  & (ID: M11BGT01; $\lambda$ =850 $\mu$m; resolution $\sim$14$''$.4; PI: Wayne S. Holland)   \\
$JCMT$ Science Archive ($^{12}$CO(J =3$-$2) & \citep[345GHz; resolution $\sim$14$''$; PROJECT = `M08BH15';][]{2009MNRAS.399.1026B}\\
\textcolor{red} {\bf Radio} \\
NRAO VLA Sky Survey  & \citep[NVSS; $\lambda$ =21 cm; resolution $\sim$46$''$;][]{1998AJ....115.1693C} \\
\hline
\end{tabular}
\end{table*}

\section{Identifying young stellar objects}
\begin{figure*}
\centering\includegraphics[width=2\columnwidth]{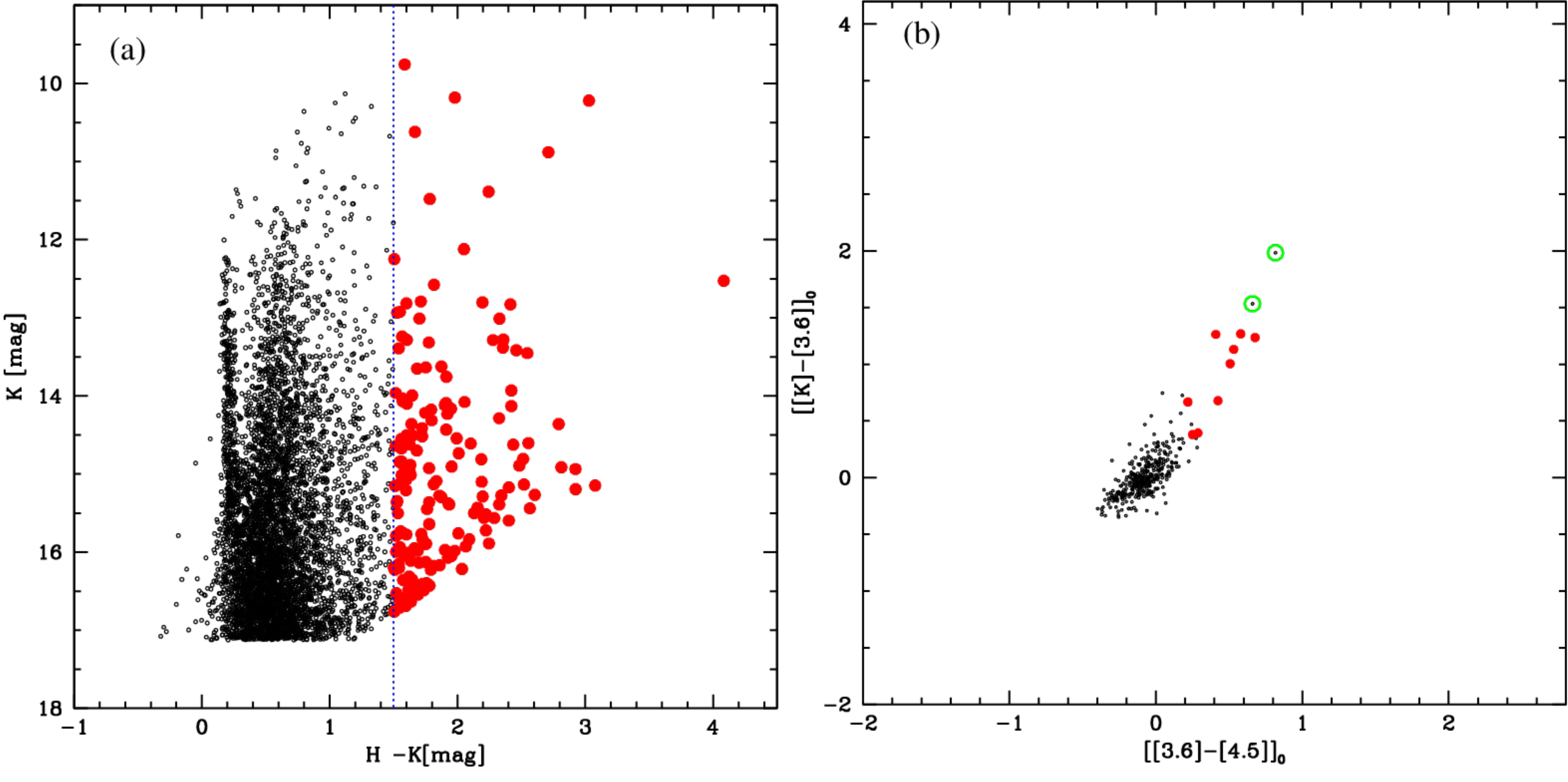}
\caption{\label{yso} (a) We show NIR CMD (H-K/K) of the sources detected in H and K bands, a cutoff value of H-K=1.5 was obtained using a nearby field region. The probable YSOs are shown with red cicles, a vertical dashed line marks the cutoff value. (b) ${[[3.6] - [4.5]]_0}$ vs. ${[[K]-[3.6]]_0}$ TCD of all the sources inside the S61 region. We used the color criteria by \citet{2009ApJS..184...18G} to identify and classify the YSOs. We show the classified Class\,{\sc i} and Class\,{\sc ii} YSOs with green and red circles, respectively.}
\end{figure*}

We have used UKIDSS NIR data along with Spitzer mid-infrared (MIR) data to identify the YSOs in the S61 region. 
Figure \ref{yso} (a) shows the H-K vs. K NIR color-magnitude diagram (CMD) made using the UKIDSS NIR data. The YSOs in the NIR CMD are found to be redder than the field stars due to their circumstellar material. In this scheme, stars above a certain cutoff value are designated probable YSOs \citep{2018ApJ...852..119B}. Using the H-K vs. K  CMD of a nearby filed ($\alpha_{2000}$: 18$^{h}$32$^{m}$28$^{s}$, $\delta_{J2000}$: -4$\degree$58$^{\prime}$48$^{\prime \prime}$) we estimated the cut off value as 1.5. The red dots in Figure \ref{yso}a show the probable YSOs identified using the UKIDSS data; the vertical dashed line is marking the cutoff value. Figure \ref{yso} (b) shows the ${[[3.4] - [4.5]]}$ vs. ${[H - K_s]}$ TCD, made using the Spitzer data, we adopted the method by \citet{2009ApJS..184...18G} to identify and classify the YSOs.

\section*{Acknowledgments}
We thank the anonymous reviewer for the comment and suggestions that were very useful and important in improving the manuscript. This work has made use of data from the European Space Agency (ESA) mission
{\it Gaia} (\footnote{https://www.cosmos.esa.int/gaia}), processed by the {\it Gaia}
Data Processing and Analysis Consortium (DPAC,
\footnote{https://www.cosmos.esa.int/web/gaia/dpac/consortium}). Funding for the DPAC
has been provided by national institutions, in particular the institutions
participating in the {\it Gaia} Multilateral Agreement.
SS acknowledge the support of the Department of Science  and Technology,  Government of India, under project No. DST/INT/Thai/P-15/2019.

\vspace{-1em}

\bibliography{61-astroph}{}

\begin{thebibliography}{}
\expandafter\ifx\csname natexlab\endcsname\relax\def\natexlab#1{#1}\fi

\bibitem[{{Anderson} {$et~al$.}(2014){Anderson}, {Bania}, {Balser},
  {Cunningham}, {Wenger}, {Johnstone}, \& {Armentrout}}]{2014ApJS..212....1A}
{Anderson}, L.~D., {Bania}, T.~M., {Balser}, D.~S., {$et~al$.} 2014, \apjs,
  212, 1

\bibitem[{{Avedisova}(2002)}]{2002ARep...46..193A}
{Avedisova}, V.~S. 2002, Astronomy Reports, 46, 193

\bibitem[{{Bailer-Jones} {$et~al$.}(2021){Bailer-Jones}, {Rybizki},
  {Fouesneau}, {Demleitner}, \& {Andrae}}]{2021AJ....161..147B}
{Bailer-Jones}, C.~A.~L., {Rybizki}, J., {Fouesneau}, M., {Demleitner}, M., \&
  {Andrae}, R. 2021, \aj, 161, 147

\bibitem[{{Baug} {$et~al$.}(2018){Baug}, {Dewangan}, {Ojha}, {Tachihara},
  {Pandey}, {Sharma}, {Tamura}, {Ninan}, \& {Ghosh}}]{2018ApJ...852..119B}
{Baug}, T., {Dewangan}, L.~K., {Ojha}, D.~K., {$et~al$.} 2018, \apj, 852, 119

\bibitem[{{Benjamin} {$et~al$.}(2005){Benjamin}, {Churchwell}, {Babler},
  {Indebetouw}, {Meade}, {Whitney}, {Watson}, {Wolfire}, {Wolff}, {Ignace},
  {Bania}, {Bracker}, {Clemens}, {Chomiuk}, {Cohen}, {Dickey}, {Jackson},
  {Kobulnicky}, {Mercer}, {Mathis}, {Stolovy}, \&
  {Uzpen}}]{2005ApJ...630L.149B}
{Benjamin}, R.~A., {Churchwell}, E., {Babler}, B.~L., {$et~al$.} 2005, \apjl,
  630, L149

\bibitem[{{Bisbas} {$et~al$.}(2009){Bisbas}, {W{\"u}nsch}, {Whitworth}, \&
  {Hubber}}]{2009A&A...497..649B}
{Bisbas}, T.~G., {W{\"u}nsch}, R., {Whitworth}, A.~P., \& {Hubber}, D.~A. 2009,
  \aap, 497, 649

\bibitem[{{Brand} {$et~al$.}(2011){Brand}, {Massi}, {Zavagno}, {Deharveng}, \&
  {Lefloch}}]{2011A&A...527A..62B}
{Brand}, J., {Massi}, F., {Zavagno}, A., {Deharveng}, L., \& {Lefloch}, B.
  2011, \aap, 527, A62

\bibitem[{{Bressert} {$et~al$.}(2012){Bressert}, {Ginsburg}, {Bally},
  {Battersby}, {Longmore}, \& {Testi}}]{2012ApJ...758L..28B}
{Bressert}, E., {Ginsburg}, A., {Bally}, J., {$et~al$.} 2012, \apjl, 758, L28

\bibitem[{{Buckle} {$et~al$.}(2009){Buckle}, {Hills}, {Smith}, {Dent}, {Bell},
  {Curtis}, {Dace}, {Gibson}, {Graves}, {Leech}, {Richer}, {Williamson},
  {Withington}, {Yassin}, {Bennett}, {Hastings}, {Laidlaw}, {Lightfoot},
  {Burgess}, {Dewdney}, {Hovey}, {Willis}, {Redman}, {Wooff}, {Berry},
  {Cavanagh}, {Davis}, {Dempsey}, {Friberg}, {Jenness}, {Kackley}, {Rees},
  {Tilanus}, {Walther}, {Zwart}, {Klapwijk}, {Kroug}, \&
  {Zijlstra}}]{2009MNRAS.399.1026B}
{Buckle}, J.~V., {Hills}, R.~E., {Smith}, H., {$et~al$.} 2009, \mnras, 399,
  1026

\bibitem[{{Churchwell} {$et~al$.}(2006){Churchwell}, {Povich}, {Allen},
  {Taylor}, {Meade}, {Babler}, {Indebetouw}, {Watson}, {Whitney}, {Wolfire},
  {Bania}, {Benjamin}, {Clemens}, {Cohen}, {Cyganowski}, {Jackson},
  {Kobulnicky}, {Mathis}, {Mercer}, {Stolovy}, {Uzpen}, {Watson}, \&
  {Wolff}}]{2006ApJ...649..759C}
{Churchwell}, E., {Povich}, M.~S., {Allen}, D., {$et~al$.} 2006, \apj, 649, 759

\bibitem[{{Condon} {$et~al$.}(1998){Condon}, {Cotton}, {Greisen}, {Yin},
  {Perley}, {Taylor}, \& {Broderick}}]{1998AJ....115.1693C}
{Condon}, J.~J., {Cotton}, W.~D., {Greisen}, E.~W., {$et~al$.} 1998, \aj, 115,
  1693

\bibitem[{{Cooper} {$et~al$.}(2013){Cooper}, {Lumsden}, {Oudmaijer}, {Hoare},
  {Clarke}, {Urquhart}, {Mottram}, {Moore}, \& {Davies}}]{2013MNRAS.430.1125C}
{Cooper}, H.~D.~B., {Lumsden}, S.~L., {Oudmaijer}, R.~D., {$et~al$.} 2013,
  \mnras, 430, 1125

\bibitem[{{Dale} {$et~al$.}(2015){Dale}, {Haworth}, \&
  {Bressert}}]{2015MNRAS.450.1199D}
{Dale}, J.~E., {Haworth}, T.~J., \& {Bressert}, E. 2015, \mnras, 450, 1199

\bibitem[{{Deharveng} {$et~al$.}(2008){Deharveng}, {Lefloch}, {Kurtz},
  {Nadeau}, {Pomar{\`e}s}, {Caplan}, \& {Zavagno}}]{2008A&A...482..585D}
{Deharveng}, L., {Lefloch}, B., {Kurtz}, S., {$et~al$.} 2008, \aap, 482, 585

\bibitem[{{Deharveng} {$et~al$.}(2005){Deharveng}, {Zavagno}, \&
  {Caplan}}]{2005A&A...433..565D}
{Deharveng}, L., {Zavagno}, A., \& {Caplan}, J. 2005, \aap, 433, 565

\bibitem[{{Dewangan} {$et~al$.}(2017{\natexlab{a}}){Dewangan}, {Ojha}, \&
  {Zinchenko}}]{2017ApJ...851..140D}
{Dewangan}, L.~K., {Ojha}, D.~K., \& {Zinchenko}, I. 2017{\natexlab{a}}, \apj,
  851, 140

\bibitem[{{Dewangan} {$et~al$.}(2017{\natexlab{b}}){Dewangan}, {Ojha},
  {Zinchenko}, {Janardhan}, \& {Luna}}]{2017ApJ...834...22D}
{Dewangan}, L.~K., {Ojha}, D.~K., {Zinchenko}, I., {Janardhan}, P., \& {Luna},
  A. 2017{\natexlab{b}}, \apj, 834, 22

\bibitem[{{Dyson} \& {Williams}(1980{\natexlab{a}})}]{1980Natur.287..373D}
{Dyson}, J.~E., \& {Williams}, D.~A. 1980{\natexlab{a}}, \nat, 287, 373

\bibitem[{{Dyson} \& {Williams}(1980{\natexlab{b}})}]{1980pim..book.....D}
---. 1980{\natexlab{b}}, {Physics of the interstellar medium}

\bibitem[{{Evans} {$et~al$.}(2009){Evans}, {Dunham}, {J{\o}rgensen}, {Enoch},
  {Mer{\'{\i}}n}, {van Dishoeck}, {Alcal{\'a}}, {Myers}, {Stapelfeldt},
  {Huard}, {Allen}, {Harvey}, {van Kempen}, {Blake}, {Koerner}, {Mundy},
  {Padgett}, \& {Sargent}}]{2009ApJS..181..321E}
{Evans}, II, N.~J., {Dunham}, M.~M., {J{\o}rgensen}, J.~K., {$et~al$.} 2009,
  \apjs, 181, 321

\bibitem[{{Fich} \& {Blitz}(1984)}]{1984ApJ...279..125F}
{Fich}, M., \& {Blitz}, L. 1984, \apj, 279, 125

\bibitem[{{Frew} {$et~al$.}(2013){Frew}, {Boji{\v{c}}i{\'c}}, \&
  {Parker}}]{2013MNRAS.431....2F}
{Frew}, D.~J., {Boji{\v{c}}i{\'c}}, I.~S., \& {Parker}, Q.~A. 2013, \mnras,
  431, 2

\bibitem[{{Fukui} {$et~al$.}(2021){Fukui}, {Inoue}, {Hayakawa}, \&
  {Torii}}]{2021PASJ...73S.405F}
{Fukui}, Y., {Inoue}, T., {Hayakawa}, T., \& {Torii}, K. 2021, \pasj, 73, S405

\bibitem[{{Gaia Collaboration} {$et~al$.}(2021){Gaia Collaboration}, {Brown},
  {Vallenari}, {Prusti}, {de Bruijne}, {Babusiaux}, {Biermann}, {Creevey},
  {Evans}, {Eyer}, {Hutton}, {Jansen}, {Jordi}, {Klioner}, {Lammers},
  {Lindegren}, {Luri}, {Mignard}, {Panem}, {Pourbaix}, {Randich}, {Sartoretti},
  {Soubiran}, {Walton}, {Arenou}, {Bailer-Jones}, {Bastian}, {Cropper},
  {Drimmel}, {Katz}, {Lattanzi}, {van Leeuwen}, {Bakker}, {Cacciari},
  {Casta{\~n}eda}, {De Angeli}, {Ducourant}, {Fabricius}, {Fouesneau},
  {Fr{\'e}mat}, {Guerra}, {Guerrier}, {Guiraud}, {Jean-Antoine Piccolo},
  {Masana}, {Messineo}, {Mowlavi}, {Nicolas}, {Nienartowicz}, {Pailler},
  {Panuzzo}, {Riclet}, {Roux}, {Seabroke}, {Sordo}, {Tanga}, {Th{\'e}venin},
  {Gracia-Abril}, {Portell}, {Teyssier}, {Altmann}, {Andrae}, {Bellas-Velidis},
  {Benson}, {Berthier}, {Blomme}, {Brugaletta}, {Burgess}, {Busso}, {Carry},
  {Cellino}, {Cheek}, {Clementini}, {Damerdji}, {Davidson}, {Delchambre},
  {Dell'Oro}, {Fern{\'a}ndez-Hern{\'a}ndez}, {Galluccio}, {Garc{\'\i}a-Lario},
  {Garcia-Reinaldos}, {Gonz{\'a}lez-N{\'u}{\~n}ez}, {Gosset}, {Haigron},
  {Halbwachs}, {Hambly}, {Harrison}, {Hatzidimitriou}, {Heiter},
  {Hern{\'a}ndez}, {Hestroffer}, {Hodgkin}, {Holl}, {Jan{\ss}en}, {Jevardat de
  Fombelle}, {Jordan}, {Krone-Martins}, {Lanzafame}, {L{\"o}ffler}, {Lorca},
  {Manteiga}, {Marchal}, {Marrese}, {Moitinho}, {Mora}, {Muinonen}, {Osborne},
  {Pancino}, {Pauwels}, {Petit}, {Recio-Blanco}, {Richards}, {Riello},
  {Rimoldini}, {Robin}, {Roegiers}, {Rybizki}, {Sarro}, {Siopis}, {Smith},
  {Sozzetti}, {Ulla}, {Utrilla}, {van Leeuwen}, {van Reeven}, {Abbas}, {Abreu
  Aramburu}, {Accart}, {Aerts}, {Aguado}, {Ajaj}, {Altavilla}, {{\'A}lvarez},
  {{\'A}lvarez Cid-Fuentes}, {Alves}, {Anderson}, {Anglada Varela}, {Antoja},
  {Audard}, {Baines}, {Baker}, {Balaguer-N{\'u}{\~n}ez}, {Balbinot}, {Balog},
  {Barache}, {Barbato}, {Barros}, {Barstow}, {Bartolom{\'e}}, {Bassilana},
  {Bauchet}, {Baudesson-Stella}, {Becciani}, {Bellazzini}, {Bernet}, {Bertone},
  {Bianchi}, {Blanco-Cuaresma}, {Boch}, {Bombrun}, {Bossini}, {Bouquillon},
  {Bragaglia}, {Bramante}, {Breedt}, {Bressan}, {Brouillet}, {Bucciarelli},
  {Burlacu}, {Busonero}, {Butkevich}, {Buzzi}, {Caffau}, {Cancelliere},
  {C{\'a}novas}, {Cantat-Gaudin}, {Carballo}, {Carlucci}, {Carnerero},
  {Carrasco}, {Casamiquela}, {Castellani}, {Castro-Ginard}, {Castro Sampol},
  {Chaoul}, {Charlot}, {Chemin}, {Chiavassa}, {Cioni}, {Comoretto}, {Cooper},
  {Cornez}, {Cowell}, {Crifo}, {Crosta}, {Crowley}, {Dafonte}, {Dapergolas},
  {David}, {David}, {de Laverny}, {De Luise}, {De March}, {De Ridder}, {de
  Souza}, {de Teodoro}, {de Torres}, {del Peloso}, {del Pozo}, {Delbo},
  {Delgado}, {Delgado}, {Delisle}, {Di Matteo}, {Diakite}, {Diener},
  {Distefano}, {Dolding}, {Eappachen}, {Edvardsson}, {Enke}, {Esquej}, {Fabre},
  {Fabrizio}, {Faigler}, {Fedorets}, {Fernique}, {Fienga}, {Figueras},
  {Fouron}, {Fragkoudi}, {Fraile}, {Franke}, {Gai}, {Garabato},
  {Garcia-Gutierrez}, {Garc{\'\i}a-Torres}, {Garofalo}, {Gavras}, {Gerlach},
  {Geyer}, {Giacobbe}, {Gilmore}, {Girona}, {Giuffrida}, {Gomel}, {Gomez},
  {Gonzalez-Santamaria}, {Gonz{\'a}lez-Vidal}, {Granvik},
  {Guti{\'e}rrez-S{\'a}nchez}, {Guy}, {Hauser}, {Haywood}, {Helmi}, {Hidalgo},
  {Hilger}, {H{\l}adczuk}, {Hobbs}, {Holland}, {Huckle}, {Jasniewicz},
  {Jonker}, {Juaristi Campillo}, {Julbe}, {Karbevska}, {Kervella}, {Khanna},
  {Kochoska}, {Kontizas}, {Kordopatis}, {Korn}, {Kostrzewa-Rutkowska},
  {Kruszy{\'n}ska}, {Lambert}, {Lanza}, {Lasne}, {Le Campion}, {Le Fustec},
  {Lebreton}, {Lebzelter}, {Leccia}, {Leclerc}, {Lecoeur-Taibi}, {Liao},
  {Licata}, {Lindstr{\o}m}, {Lister}, {Livanou}, {Lobel}, {Madrero Pardo},
  {Managau}, {Mann}, {Marchant}, {Marconi}, {Marcos Santos}, {Marinoni},
  {Marocco}, {Marshall}, {Martin Polo}, {Mart{\'\i}n-Fleitas}, {Masip},
  {Massari}, {Mastrobuono-Battisti}, {Mazeh}, {McMillan}, {Messina},
  {Michalik}, {Millar}, {Mints}, {Molina}, {Molinaro}, {Moln{\'a}r},
  {Montegriffo}, {Mor}, {Morbidelli}, {Morel}, {Morris}, {Mulone}, {Munoz},
  {Muraveva}, {Murphy}, {Musella}, {Noval}, {Ord{\'e}novic}, {Orr{\`u}},
  {Osinde}, {Pagani}, {Pagano}, {Palaversa}, {Palicio}, {Panahi}, {Pawlak},
  {Pe{\~n}alosa Esteller}, {Penttil{\"a}}, {Piersimoni}, {Pineau}, {Plachy},
  {Plum}, {Poggio}, {Poretti}, {Poujoulet}, {Pr{\v{s}}a}, {Pulone}, {Racero},
  {Ragaini}, {Rainer}, {Raiteri}, {Rambaux}, {Ramos}, {Ramos-Lerate}, {Re
  Fiorentin}, {Regibo}, {Reyl{\'e}}, {Ripepi}, {Riva}, {Rixon}, {Robichon},
  {Robin}, {Roelens}, {Rohrbasser}, {Romero-G{\'o}mez}, {Rowell}, {Royer},
  {Rybicki}, {Sadowski}, {Sagrist{\`a} Sell{\'e}s}, {Sahlmann}, {Salgado},
  {Salguero}, {Samaras}, {Sanchez Gimenez}, {Sanna}, {Santove{\~n}a},
  {Sarasso}, {Schultheis}, {Sciacca}, {Segol}, {Segovia}, {S{\'e}gransan},
  {Semeux}, {Shahaf}, {Siddiqui}, {Siebert}, {Siltala}, {Slezak}, {Smart},
  {Solano}, {Solitro}, {Souami}, {Souchay}, {Spagna}, {Spoto}, {Steele},
  {Steidelm{\"u}ller}, {Stephenson}, {S{\"u}veges}, {Szabados}, {Szegedi-Elek},
  {Taris}, {Tauran}, {Taylor}, {Teixeira}, {Thuillot}, {Tonello}, {Torra},
  {Torra}, {Turon}, {Unger}, {Vaillant}, {van Dillen}, {Vanel}, {Vecchiato},
  {Viala}, {Vicente}, {Voutsinas}, {Weiler}, {Wevers}, {Wyrzykowski}, {Yoldas},
  {Yvard}, {Zhao}, {Zorec}, {Zucker}, {Zurbach}, \&
  {Zwitter}}]{2021A&A...649A...1G}
{Gaia Collaboration}, {Brown}, A.~G.~A., {Vallenari}, A., {$et~al$.} 2021,
  \aap, 649, A1

\bibitem[{{Gutermuth} {$et~al$.}(2009){Gutermuth}, {Megeath}, {Myers}, {Allen},
  {Pipher}, \& {Fazio}}]{2009ApJS..184...18G}
{Gutermuth}, R.~A., {Megeath}, S.~T., {Myers}, P.~C., {$et~al$.} 2009, \apjs,
  184, 18

\bibitem[{{Gutermuth} {$et~al$.}(2005){Gutermuth}, {Megeath}, {Pipher},
  {Williams}, {Allen}, {Myers}, \& {Raines}}]{2005ApJ...632..397G}
{Gutermuth}, R.~A., {Megeath}, S.~T., {Pipher}, J.~L., {$et~al$.} 2005, \apj,
  632, 397

\bibitem[{{Hern{\'a}ndez} {$et~al$.}(2004){Hern{\'a}ndez}, {Calvet},
  {Brice{\~n}o}, {Hartmann}, \& {Berlind}}]{2004AJ....127.1682H}
{Hern{\'a}ndez}, J., {Calvet}, N., {Brice{\~n}o}, C., {Hartmann}, L., \&
  {Berlind}, P. 2004, \aj, 127, 1682

\bibitem[{{Hunter} \& {Massey}(1990)}]{1990AJ.....99..846H}
{Hunter}, D.~A., \& {Massey}, P. 1990, \aj, 99, 846

\bibitem[{{Hunter} {$et~al$.}(1990){Hunter}, {Thronson}, \&
  {Wilton}}]{1990AJ....100.1915H}
{Hunter}, D.~A., {Thronson}, Jr., H.~A., \& {Wilton}, C. 1990, \aj, 100, 1915

\bibitem[{{Lucas} {$et~al$.}(2008){Lucas}, {Hoare}, {Longmore}, {Schr{\"o}der},
  {Davis}, {Adamson}, {Bandyopadhyay}, {de Grijs}, {Smith}, {Gosling},
  {Mitchison}, {G{\'a}sp{\'a}r}, {Coe}, {Tamura}, {Parker}, {Irwin}, {Hambly},
  {Bryant}, {Collins}, {Cross}, {Evans}, {Gonzalez-Solares}, {Hodgkin},
  {Lewis}, {Read}, {Riello}, {Sutorius}, {Lawrence}, {Drew}, {Dye}, \&
  {Thompson}}]{2008MNRAS.391..136L}
{Lucas}, P.~W., {Hoare}, M.~G., {Longmore}, A., {$et~al$.} 2008, \mnras, 391,
  136

\bibitem[{{Manoj} {$et~al$.}(2006){Manoj}, {Bhatt}, {Maheswar}, \&
  {Muneer}}]{2006ApJ...653..657M}
{Manoj}, P., {Bhatt}, H.~C., {Maheswar}, G., \& {Muneer}, S. 2006, \apj, 653,
  657

\bibitem[{{Molinari} {$et~al$.}(2010){Molinari}, {Swinyard}, {Bally}, {Barlow},
  {Bernard}, {Martin}, {Moore}, {Noriega-Crespo}, {Plume}, {Testi}, {Zavagno},
  {Abergel}, {Ali}, {Andr{\'e}}, {Baluteau}, {Benedettini}, {Bern{\'e}},
  {Billot}, {Blommaert}, {Bontemps}, {Boulanger}, {Brand}, {Brunt}, {Burton},
  {Campeggio}, {Carey}, {Caselli}, {Cesaroni}, {Cernicharo}, {Chakrabarti},
  {Chrysostomou}, {Codella}, {Cohen}, {Compiegne}, {Davis}, {de Bernardis}, {de
  Gasperis}, {Di Francesco}, {di Giorgio}, {Elia}, {Faustini}, {Fischera},
  {Fukui}, {Fuller}, {Ganga}, {Garcia-Lario}, {Giard}, {Giardino}, {Glenn},
  {Goldsmith}, {Griffin}, {Hoare}, {Huang}, {Jiang}, {Joblin}, {Joncas},
  {Juvela}, {Kirk}, {Lagache}, {Li}, {Lim}, {Lord}, {Lucas}, {Maiolo},
  {Marengo}, {Marshall}, {Masi}, {Massi}, {Matsuura}, {Meny}, {Minier},
  {Miville-Desch{\^e}nes}, {Montier}, {Motte}, {M{\"u}ller}, {Natoli}, {Neves},
  {Olmi}, {Paladini}, {Paradis}, {Pestalozzi}, {Pezzuto}, {Piacentini},
  {Pomar{\`e}s}, {Popescu}, {Reach}, {Richer}, {Ristorcelli}, {Roy}, {Royer},
  {Russeil}, {Saraceno}, {Sauvage}, {Schilke}, {Schneider-Bontemps},
  {Schuller}, {Schultz}, {Shepherd}, {Sibthorpe}, {Smith}, {Smith},
  {Spinoglio}, {Stamatellos}, {Strafella}, {Stringfellow}, {Sturm}, {Taylor},
  {Thompson}, {Tuffs}, {Umana}, {Valenziano}, {Vavrek}, {Viti}, {Waelkens},
  {Ward-Thompson}, {White}, {Wyrowski}, {Yorke}, \&
  {Zhang}}]{2010PASP..122..314M}
{Molinari}, S., {Swinyard}, B., {Bally}, J., {$et~al$.} 2010, \pasp, 122, 314

\bibitem[{{Panagia}(1973)}]{1973AJ.....78..929P}
{Panagia}, N. 1973, \aj, 78, 929

\bibitem[{{Pandey} {$et~al$.}(2020){Pandey}, {Sharma}, {Panwar}, {Dewangan},
  {Ojha}, {Bisen}, {Sinha}, {Ghosh}, \& {Pandey}}]{2020ApJ...891...81P}
{Pandey}, R., {Sharma}, S., {Panwar}, N., {$et~al$.} 2020, \apj, 891, 81

\bibitem[{{Pandey} {$et~al$.}(2022){Pandey}, {Sharma}, {Dewangan}, {Ojha},
  {Panwar}, {Das}, {Bisen}, {Ghosh}, \& {Sinha}}]{2022ApJ...926...25P}
{Pandey}, R., {Sharma}, S., {Dewangan}, L.~K., {$et~al$.} 2022, \apj, 926, 25

\bibitem[{{Parker} {$et~al$.}(2005){Parker}, {Phillipps}, {Pierce}, {Hartley},
  {Hambly}, {Read}, {MacGillivray}, {Tritton}, {Cass}, {Cannon}, {Cohen},
  {Drew}, {Frew}, {Hopewell}, {Mader}, {Malin}, {Masheder}, {Morgan}, {Morris},
  {Russeil}, {Russell}, \& {Walker}}]{2005MNRAS.362..689P}
{Parker}, Q.~A., {Phillipps}, S., {Pierce}, M.~J., {$et~al$.} 2005, \mnras,
  362, 689

\bibitem[{{Schmiedeke} {$et~al$.}(2016){Schmiedeke}, {Schilke}, {M{\"o}ller},
  {S{\'a}nchez-Monge}, {Bergin}, {Comito}, {Csengeri}, {Lis}, {Molinari},
  {Qin}, \& {Rolffs}}]{2016A&A...588A.143S}
{Schmiedeke}, A., {Schilke}, P., {M{\"o}ller}, T., {$et~al$.} 2016, \aap, 588,
  A143

\bibitem[{{Sharma} {$et~al$.}(2020){Sharma}, {Ghosh}, {Ojha}, {Pandey},
  {Sinha}, {Pandey}, {Ghosh}, {Panwar}, \& {Pandey}}]{2020MNRAS.498.2309S}
{Sharma}, S., {Ghosh}, A., {Ojha}, D.~K., {$et~al$.} 2020, \mnras, 498, 2309

\bibitem[{{Shima} {$et~al$.}(2017){Shima}, {Tasker}, \&
  {Habe}}]{2017MNRAS.467..512S}
{Shima}, K., {Tasker}, E.~J., \& {Habe}, A. 2017, \mnras, 467, 512

\bibitem[{{Stahler} \& {Palla}(2005)}]{2005fost.book.....S}
{Stahler}, S.~W., \& {Palla}, F. 2005, {The Formation of Stars}, 865

\bibitem[{{Vioque} {$et~al$.}(2018){Vioque}, {Oudmaijer}, {Baines},
  {Mendigut{\'\i}a}, \& {P{\'e}rez-Mart{\'\i}nez}}]{2018A&A...620A.128V}
{Vioque}, M., {Oudmaijer}, R.~D., {Baines}, D., {Mendigut{\'\i}a}, I., \&
  {P{\'e}rez-Mart{\'\i}nez}, R. 2018, \aap, 620, A128

\bibitem[{{Wright} {$et~al$.}(2010){Wright}, {Eisenhardt}, {Mainzer},
  {Ressler}, {Cutri}, {Jarrett}, {Kirkpatrick}, {Padgett}, {McMillan},
  {Skrutskie}, {Stanford}, {Cohen}, {Walker}, {Mather}, {Leisawitz}, {Gautier},
  {McLean}, {Benford}, {Lonsdale}, {Blain}, {Mendez}, {Irace}, {Duval}, {Liu},
  {Royer}, {Heinrichsen}, {Howard}, {Shannon}, {Kendall}, {Walsh}, {Larsen},
  {Cardon}, {Schick}, {Schwalm}, {Abid}, {Fabinsky}, {Naes}, \&
  {Tsai}}]{2010AJ....140.1868W}
{Wright}, E.~L., {Eisenhardt}, P.~R.~M., {Mainzer}, A.~K., {$et~al$.} 2010,
  \aj, 140, 1868

\bibitem[{{Yang} {$et~al$.}(2019){Yang}, {Chen}, {Shen}, {Li}, {Wang}, {Jiang},
  {Li}, {Dong}, {Wu}, \& {Qiao}}]{2019ApJS..241...18Y}
{Yang}, K., {Chen}, X., {Shen}, Z.-Q., {$et~al$.} 2019, \apjs, 241, 18

\bibitem[{{Ybarra} \& {Lada}(2009)}]{2009ApJ...695L.120Y}
{Ybarra}, J.~E., \& {Lada}, E.~A. 2009, \apjl, 695, L120

\bibitem[{{Zavagno} {$et~al$.}(2010){Zavagno}, {Russeil}, {Motte}, {Anderson},
  {Deharveng}, {Rod{\'o}n}, {Bontemps}, {Abergel}, {Baluteau}, {Sauvage},
  {Andr{\'e}}, {Hill}, \& {White}}]{2010A&A...518L..81Z}
{Zavagno}, A., {Russeil}, D., {Motte}, F., {$et~al$.} 2010, \aap, 518, L81

\end{thebibliography}
\bibliographystyle{aasjournal}

\end{document}